# Democratizing Generative AI for Sustainable Competitive Advantage

Carlos J. Costa
ISEG – Lisbon School of Economics and Management,
Universidade de Lisboa,
Lisboa, Portugal
cjcosta@iseg.ulisboa.pt

Joao Tiago Aparício
INESC - ID and Instituto Superior Técnico,
Universidade de Lisboa,
Lisboa, Portugal
joao.aparicio@tecnico.ulisboa.pt

Manuela Aparício
NOVA Information Management School (NOVA IMS),
Universidade NOVA de Lisboa,
Lisboa, Portugal
manuela.aparicio@novaims.unl.pt

***Abstract*—As generative artificial intelligence (GenAI) diffuses across industries and becomes broadly accessible, the locus of sustainable competitive advantage shifts from technology ownership toward the quality of employee-level adoption and use. This paper develops a cross-level conceptual framework linking firm-level GenAI investment and governance to individual-level AI democratization, defined as the extent to which employees meaningfully, responsibly, and effectively use GenAI in their daily work. We argue that individual-level AI democratization, grounded in three micro foundations (AI usefulness, ease of use, and AI literacy), mediates the relationship between organizational GenAI investments and sustainable competitive advantage. Drawing on the technology acceptance model, resource-based theory, and emerging empirical evidence on AI productivity effects, we advance six propositions linking perceived usefulness, ease of use, AI literacy, responsible use, and innovation outcomes to organizational transformation and sustained relative performance. The framework provides a measurement scaffold for empirical research and offers managerial guidance on treating GenAI as augmentation infrastructure rather than solely as automation. We conclude by outlining future research directions, including longitudinal and cross-cultural investigations of literacy, governance, and transformation dynamics.**

## I. Introduction

Generative artificial intelligence (GenAI) has rapidly transitioned from a specialist capability to an embedded component of routine workflows across industries (Costa et al., 2024a). Large language models, image generators, and multimodal systems are now accessible through consumer-grade interfaces, enterprise platforms, and application programming interfaces that require minimal technical expertise. This diffusion fundamentally alters the strategic calculus surrounding AI-based competitive advantage. When access to technology broadens and scarcity diminishes, the source of durable advantage shifts away from the technology itself and toward execution, integration, and human capability Several authors (e.g. [1], [2]) analyse the economic implications of generative AI demands a multidimensional approach, with various methodologies, such as agent-based modelling, econometric models, input-output analysis, reinforcement learning, surveys, scenario analysis, policy analysis, each offering distinct strengths and limitations in terms of uncertainty handling and resource requirements. Selecting the appropriate combination of these methods, based on specific research goals and context, enables a comprehensive understanding of how generative AI affects the economy. Despite substantial organizational investment in GenAI, evidence on productivity outcomes remains ambiguous. Some studies report significant productivity gains at the task level ([3], [4]), while other studies report uneven adoption, misuse risks, and unrealized potential at the organizational level [5]. This ambiguity suggests that the missing link lies not in whether organizations adopt GenAI, but in how deeply, broadly, and responsibly employees appropriate these tools in their daily work.

This paper addresses the micro foundation gap by developing a conceptual framework centered on individual - level AI democratization, the extent to which employees meaningfully, responsibly, and effectively use GenAI in their daily work. We argue that this construct mediates the relationship between firm-level GenAI investment and sustainable competitive advantage. Our central research question (RQ) is:

RQ: How does the democratization of generative AI at the individual level contribute to sustainable competitive advantage in organizations?

We contribute to the literature in three ways. First, we shift the analytical lens from firm-level adoption to employee-level value creation, addressing calls for micro-level theorizing in information systems research [6]. Second, we build a cross-level conceptual model with explicit micro foundations and testable propositions that link individual perceptions, behaviors, and outcomes to organizational transformation. Third, we provide a measurement scaffold for empirical work across industries and over time, enabling researchers to operationalize constructs at multiple levels of analysis.

The remainder of this paper proceeds as follows; Section 2 reviews the theoretical background on technology acceptance, AI productivity, and sustainable competitive advantage; Section 3 presents the conceptual framework and defines key constructs; Section 4 develops six propositions linking individual level AI democratization to organizational outcomes; Section 5 discusses operationalization for empirical testing; Section 6 addresses managerial implications; and Section 7 concludes with contributions and future research directions..

## II. Theoretical Background

### *A. The Shifting Locus of AI - Based Advantage*

Traditional perspectives on information technology and competitive advantage emphasize resource heterogeneity and

This paper was presented at the BIG AI @ MIT, 2026

imperfect mobility [7]. Firms that possess rare, valuable, and difficult-to-imitate IT capabilities can sustain superior performance relative to competitors. However, the rapid commoditization of GenAI challenges this logic. When foundational models are available through commercial APIs, when open-source alternatives proliferate, and when user interfaces require no programming expertise, technology access alone cannot constitute a source of sustained differentiation.

Davenport and Ronanki anticipated this shift, arguing that AI's business value depends less on algorithmic sophistication than on integration with business processes, data quality, and human judgment [1]. As GenAI diffuses, the argument is not that technology stops mattering; rather, firm value increasingly depends on how broadly and responsibly employees appropriate GenAI capabilities [8].

This perspective aligns with the resource-based view's emphasis on complementary assets and organizational capabilities [6]. GenAI tools are necessary but insufficient for competitive advantage; their value is realized through workforce capability, governance structures, and workflow integration that enable effective appropriation.

### B. *Technology Acceptance and Use*

The technology acceptance theory and its extensions provide foundational constructs for understanding individual-level technology adoption [5]. Perceived usefulness, the degree to which a person believes that using a technology will enhance job performance, and perceived ease of use, the degree to which a person believes that using a technology will be free of effort, are established predictors of behavioral intention and actual use.

In the GenAI context, these constructs take on particular significance. Perceived usefulness reflects whether employees believe GenAI tools genuinely improve their productivity, creativity, or decision quality. Perceived ease of use captures the cognitive friction associated with prompt engineering, output verification, and workflow integration. Both constructs influence whether employees move from initial trial to sustained, meaningful use.

However, acceptance theory was developed for technologies with deterministic outputs. GenAI introduces probabilistic, sometimes unreliable outputs that require verification and judgment. This characteristic elevates the importance of AI literacy, the knowledge and skills required to critically evaluate, appropriately apply, and responsibly use AI-generated outputs [9]. AI literacy extends beyond technical proficiency to encompass understanding of AI limitations, bias risks, and ethical considerations.

### C. *AI Literacy and Responsible Use*

Ng and colleagues conceptualize AI literacy as encompassing four dimensions: knowing and understanding AI; using and applying AI; evaluating and creating AI; and AI ethics [9]. For GenAI specifically, literacy involves understanding how large language models generate outputs, recognizing the risks of hallucination, applying appropriate verification strategies, and making ethical judgments about use contexts.

Responsible use emerges as a critical construct distinct from mere frequency of use. Venkatesh and colleagues distinguish between different patterns of technology use [5]; in the GenAI context, responsible use involves verification behaviors, appropriate task selection, bias awareness, and ethical compliance. Organizations face risks not only from underuse (failing to capture productivity benefits) but also from misuse (overreliance, bias propagation, and ethical violations). The European Union's AI Act [10] codifies responsible use requirements, creating regulatory pressure for organizations to ensure employees use AI systems appropriately. This regulatory context reinforces the strategic importance of AI literacy as a foundation for responsible use.

### D. *From Individual Use to Organizational Outcomes*

The link between individual-level technology use and organizational performance has been theorized through multiple mechanisms. Melville and colleagues propose an integrative model in which IT resources generate business value through business processes, subject to competitive and macro-environmental factors [6]. Individual productivity gains aggregate to process improvements, which contribute to organizational performance.

For GenAI, this aggregation mechanism operates through several channels. Task-level productivity gains, documented in experimental studies showing 30% to 40% improvements in writing and coding tasks [3], can accumulate across employees and tasks. Innovation effects emerge when employees use GenAI for ideation, experimentation, and idea recombination. Quality improvements occur when GenAI assists with analysis, error detection, and decision support.

However, aggregation is not automatic. Organizational transformation in the age of generative AI goes far beyond simply adopting new tools; it requires a fundamental rethinking of how work is structured, executed, and continuously improved. This involves redesigning core business processes so that human effort and AI capabilities are meaningfully integrated, redefining roles to reflect collaboration between people and intelligent systems, and building new capabilities across the organization, such as data literacy, prompt engineering, and AI governance. As Davenport and Ronanki suggest, the real value of AI is unlocked not through isolated use cases, but through systemic changes that embed AI into everyday workflows and decision-making[1].

Importantly, sustainable competitive advantage does not arise merely from access to generative AI technologies, which are increasingly widespread and accessible. Instead, it emerges from how effectively an organization transforms itself to use these tools at scale. Companies that succeed are those that develop the internal ability to democratize AI, making it usable, understandable, and valuable across functions and levels of the organization. This includes creating supportive cultures, investing in training, aligning incentives, and establishing governance structures that encourage responsible experimentation and innovation. What makes this advantage difficult for competitors to replicate is not the technology itself, but the complex combination of organizational elements required to support it. These include tacit knowledge, cross-functional collaboration, leadership commitment, and embedded routines that evolve over time. In this sense, the capability to operationalize AI becomes a form of organizational capital deeply ingrained, path-dependent, and resistant to imitation. As a result, firms that successfully align their structures, people, and processes around AI are better positioned to achieve lasting performance gains and maintain a competitive edge in rapidly changing environments.

## III. CONCEPTUAL FRAMEWORK

### A. Framework Overview

The conceptual framework proposed in this study explains the main determinants of sustainable competitiveness advantage funded in information systems success theory [11], firm investments in technology, capability development, mediated by the level of AI democratization and AI Literacy of users [9], by the inducing of a systemic transformation and understanding the impacts in productivity, technology responsible usage on the sustainability of organizations. Figure 1 presents our conceptual framework linking firm-level GenAI investment and governance to sustainable competitive advantage through individual-level AI democratization. The framework operates across three levels: organizational (inputs and macro-outcomes), adoption and workflow (access conditions and appropriation quality), and individual (employee perceptions, skills, use patterns, and outcomes).

At the individual level, we distinguish between antecedents, behaviors, and outcomes. Perceived AI usefulness (AI_U) and ease of use (AI_EOU) are perceptual antecedents drawn from technology acceptance theory. AI literacy (AI_Lit) represents the foundation of knowledge and skills. Actual GenAI usage (USE) captures the frequency and breadth of behavioral use. Responsible use (RUSE_I) reflects individual level verification, compliance, and ethical behavior. Productivity and task performance (PROD) and individual innovation (INNOV_I) are proximate outcomes.

These individual level constructs aggregate into organizational outcomes: organizational transformation (TRANSF), encompassing process redesign, role evolution, and capability development; and sustainable competitive advantage (SCA), reflecting sustained relative performance, agility, and differentiation.

### B. Construct Definitions

**CONCEPTUAL FRAMEWORK OF INDIVIDUAL-LEVEL AI DEMOCRATIZATION FOR SUSTAINABLE COMPETITIVE ADVANTAGE**

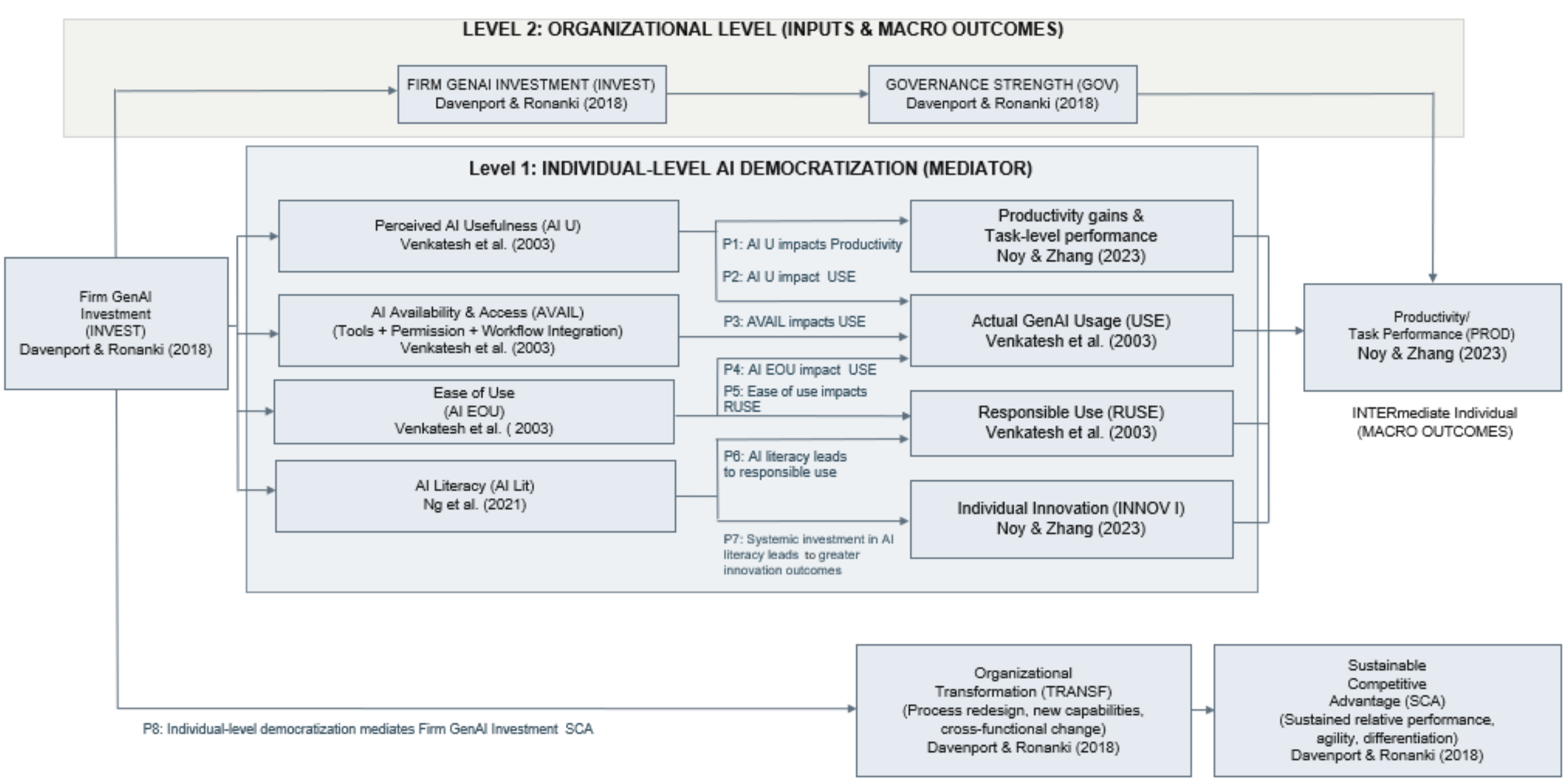

Fig. 1. Conceptual Framework of Individual Level AI Democratization for Sustainable Competitive Advantage

At the organizational level, firm GenAI investment (INVEST) encompasses technology acquisition, infrastructure development, training programs, and change management initiatives. Governance strength (GOV) captures policy clarity, acceptable - use guidelines, monitoring mechanisms, and ethical frameworks. These organizational inputs shape the conditions under which employees encounter and adopt GenAI tools.[12]

At the adoption and workflow level, availability (AVAIL) reflects access conditions of what employees can use, which tools, in which workflow steps, and through which interfaces. Responsible use environment (RUSE) captures the quality of actual appropriation, including verification norms, bias awareness, and ethical compliance at the workflow level.

Individual level AI democratization is defined as the extent to which employees meaningfully, responsibly, and effectively use GenAI in their daily work [13]. This meta-construct encompasses three micro foundations: AI usefulness, Ease of use, and AI literacy.

AI usefulness is the degree to which an employee perceives that GenAI tools enhance their job performance, productivity, or work quality.

Ease of use is the degree to which an employee perceives that using GenAI tools is free of cognitive effort, including prompt formulation, output interpretation, and workflow integration.

AI literacy is the knowledge, skills, and critical understanding required to evaluate AI outputs, recognize limitations, apply appropriate verification strategies, and make ethical judgments about the use of GenAI.[14]

Responsible use is defined as GenAI usage patterns characterized by appropriate task selection, output verification, bias awareness, and ethical compliance. Responsible use is distinguished from mere frequency; high-frequency use without verification constitutes overreliance rather than responsible use.[14]

Sustainable competitive advantage is defined as sustained superior performance relative to competitors, achieved through organizational capabilities that are valuable, rare, and difficult to imitate. In the GenAI context, SCA derives not from technology ownership but from the organizational capability to democratize AI effectively across the workforce.

### C. *Cross-Level Mechanisms*

The framework specifies cross-level mechanisms linking organizational inputs to individual behaviors and aggregating individual outcomes into organizational performance. Cross-level mechanisms have been studied in various contexts, mostly based on information systems success theory [11], [15]. Those mechanisms may be grouped into: Top-down mechanisms, Bottom-up mechanisms, and Mediation.

Firm investment shapes availability, which, in turn, influences perceived usefulness and ease of use. Governance strength shapes the responsible-use environment, which in turn influences AI literacy development and responsible-use behaviors. Training investments directly affect AI literacy.

Individual productivity gains aggregate through workflow processes into organizational productivity. Individual innovation contributes to organizational innovation capability. Patterns of responsible use aggregate to organizational risk profiles and compliance postures.

Individual-level AI democratization mediates the relationship between firm GenAI investment and sustainable competitive advantage. Organizations that invest heavily but fail to achieve broad, responsible employee adoption will not realize competitive benefits. Conversely, organizations that achieve high-quality democratization can gain an advantage even with moderate investment levels.

## IV. PROPOSITIONS

Drawing on the conceptual framework and theoretical background, we advance six propositions linking individual-level constructs to use behaviors, responsible use, innovation, and organizational outcomes.

### A. *Proposition 1: AI Usefulness Impacts Productivity and Use*

Technology acceptance research consistently demonstrates that perceived usefulness is the strongest predictor of technology adoption and continued use [5]. When employees believe that GenAI tools genuinely enhance their job performance, they are more likely to invest effort in learning to use these tools effectively and to integrate them into routine workflows. Experimental evidence supports this relationship in the GenAI context. Noy and Zhang found that workers who perceived ChatGPT as useful for their writing tasks showed higher adoption rates and larger productivity gains [3]. The usefulness perception reflects an assessment of whether technology delivers value that outweighs the effort of adoption.

Proposition 1: Higher perceived AI usefulness is positively associated with (a) productivity gains, (b) task - level performance improvements, and (c) actual GenAI use.

### B. ***Proposition 2: Ease of Use Impacts Actual Usage***

Perceived ease of use influences adoption both directly and indirectly through perceived usefulness (Venkatesh et al., 2003). When GenAI tools are perceived as difficult to use requiring complex prompt engineering, producing outputs that are hard to interpret, or integrating poorly with existing workflows, employees face cognitive friction that discourages sustained use. The ease-of-use construct is particularly relevant for GenAI [16], as effective use requires skills that many employees lack. Formulating effective prompts, iterating on outputs, and integrating AI-generated content into work products all require effort. Tools and interfaces that reduce this friction, through templates, examples, or workflow integration, can increase actual usage.

Proposition 2: Higher perceived ease of use is positively associated with actual GenAI usage.

### C. ***Proposition 3: Ease of Use Impacts Perceived Usefulness***

Ease of use also influences perceived usefulness. When technology is difficult to use, employees may underestimate its potential value because they cannot effectively access its capabilities. Conversely, when technology is easy to use, employees can more readily experience its benefits, thereby reinforcing perceptions of usefulness [16]. This relationship suggests that organizations seeking to increase GenAI adoption should attend not only to making tools available but also to reducing friction in their use. Training, interface design, and workflow integration all contribute to ease of use, which, in turn, shapes perceptions of usefulness.

Proposition 3: Higher perceived ease of use is positively associated with perceived AI usefulness.

### D. ***Proposition 4: AI Literacy Enables Responsible Use***

AI literacy provides the foundation for responsible GenAI use. Employees who understand how large language models work, recognize the risks of hallucination, and possess verification skills are better equipped to use GenAI appropriately. Conversely, employees who lack AI literacy may over-rely on AI outputs, fail to detect errors, or use GenAI in inappropriate contexts. Ng and colleagues argue that AI literacy encompasses not only technical understanding but also ethical reasoning [9]. Literate users can make informed judgments about when GenAI use is appropriate, what verification is required, and how to avoid bias propagation. This literacy foundation enables responsible use patterns that reduce organizational risk.

Proposition 4: Higher AI literacy is positively associated with responsible GenAI use, characterized by appropriate verification behaviors, bias awareness, and reduced overreliance.

### E. ***Proposition 5: Literacy Investment Drives Innovation***

Organizations that invest systematically in AI literacy create conditions for innovation beyond productivity gains. Literate employees can explore novel applications, combine AI capabilities with domain expertise, and engage in human-AI recombination to generate new ideas and approaches. Brynjolfsson and colleagues document that AI-assisted workers not only completed tasks faster but also explored more solution approaches [4]. This exploration behavior, enabled by literacy, contributes to individual and organizational innovation. Organizations that treat AI literacy as a strategic investment rather than a compliance requirement may realize innovation benefits that competitors cannot easily replicate.

Proposition 5: Organizations that systematically invest in AI literacy achieve stronger innovation outcomes through human-AI recombination and exploration.

### F. *Proposition 6: Democratization Mediates Investment - Advantage Relationship*

The central proposition of our framework is that individual-level AI democratization mediates the relationship between firm GenAI investment and sustainable competitive advantage. Investment alone is insufficient; advantage accrues to organizations that translate investment into broad, responsible, effective employee use. This mediation mechanism explains why organizations with similar investment levels may achieve different outcomes. Organizations that achieve high-quality democratization, characterized by widespread adoption, responsible use patterns, and literacy-enabled innovation, can sustain advantage even as technology diffuses. The organizational capability to democratize effectively becomes the source of advantage, not the technology itself [8].

Proposition 6: Individual-level AI democratization mediates the relationship between firm GenAI investment and sustainable competitive advantage. Organizations that achieve higher levels of democratization, characterized by broad adoption, responsible use, and literacy-enabled innovation, lead to greater sustained competitive benefits from their GenAI investments.

## V. Conceptual Repositioning: Where Does Sustainable Advantage Come From?

### A. *The Diffusion Trajectory*

Understanding the origins of sustainable advantage requires examining the trajectory of GenAI diffusion. We identify three phases that reshape the competitive landscape: Phase 1: Scarce access to technology. In the early stages, competitive advantage derives from technology ownership. Organizations with access to advanced AI capabilities through proprietary development, early partnerships, or substantial infrastructure investment can outperform competitors who lack such access. This phase characterized the pre-2022 period, when large language models required significant computational resources and expertise. In Phase 2: Broad tool availability. As foundational models become available through commercial APIs and consumer interfaces, technology access ceases to be a differentiator. The release of ChatGPT in November 2022 marked a transition point; within months, GenAI tools became accessible to virtually any organization and individual. In this phase, the advantage shifts from having the technology to deploying it effectively. In Phase 3: Differentiation in workforce capability and use quality. Under broad diffusion, firm value increasingly depends on how extensively and responsibly employees appropriate GenAI capabilities. Organizations differentiate not through technology but through the depth, breadth, and quality of employee-level adoption. This phase, which we argue characterizes the current competitive environment, elevates the importance of individual-level AI democratization.

### B. *The Democratization Imperative*

The argument is not that technology stops mattering; rather, technology becomes necessary but insufficient. All competitors can access similar GenAI capabilities; differentiation emerges from organizational capabilities that shape how those technologies are used [8].

This repositioning has profound implications for strategy. Organizations cannot sustain an advantage through technology acquisition alone. Instead, they must build capabilities in: Workforce development, consists of Cultivating AI literacy across the employee base, not just among technical specialists; Governance design addresses creating frameworks that enable experimentation while managing risks; Workflow integration corresponds to the embedding GenAI into business processes in ways that capture value; Culture shaping are the fostering norms of responsible use, verification, and continuous learning.[12]

These capabilities are organizationally embedded, path-dependent, and difficult to imitate, precisely the characteristics that resource-based theory identifies as sources of sustainable advantage [7].

## VI. Operationalization for Empirical Testing

### A. *Multi-Level Measurement Approach*

Empirical testing of the framework requires measurement at multiple levels of analysis. Table 1 presents illustrative indicators for each construct, organized by level.

TABLE I. Operationalization of Constructs

| Level | Construct | Illustrative Indicators |
|---|---|---|
| Firm | INVEST / GOV | Rollout intensity, workflow integration, policy clarity, and monitoring strength |
| Workflow | AVAIL | Employee access rights, sanctioned tools, and embedded workflow touchpoints |
| Individual | AI_U / AI_EOU / AI_Lit | Perceived usefulness, cognitive friction, capability, risk understanding |
| Use quality | USE / RUSE | Frequency, breadth, verification behavior, bias checks, compliance |
| Outcomes | PROD / INNOV_I / TRANSF / SCA | | Task quality, ideation, redesign, agility, performance relative to peers |

### B. *Firm-Level Measures*

GenAI Investment (INVEST) can be operationalized through technology expenditure (licenses, API costs, infrastructure), training investment (hours, programs, coverage), change management resources (dedicated personnel, communication intensity), and integration effort (workflow redesign initiatives, pilot programs).

Governance Strength (GOV) can be assessed through policy existence and clarity (acceptable use policies, ethical guidelines), monitoring mechanisms (usage tracking, output auditing), enforcement practices (violation consequences, compliance verification), and governance maturity (ad hoc vs. systematic approaches).

### C. *Workflow-Level Measures*

Availability (AVAIL) captures access conditions, such as tool access breadth (which employees can access which tools), workflow integration depth (embedded vs. standalone access), use case sanctioning (approved vs. prohibited applications), and Interface quality (enterprise platforms vs. consumer tools).

Responsible Use Environment (RUSE) at the workflow level can be operationalized through verification norms (expected checking behaviors), quality standards (output review requirements), escalation procedures (handling uncertain or sensitive cases), and peer accountability (team-level responsibility for appropriate use).

### D. *Individual-Level Measures*

Perceived AI Usefulness (AI_U) can be measured through adapted TAM scales: "Using GenAI tools improves my job performance", "GenAI tools increase my productivity", "GenAI tools enhance the quality of my work", "GenAI tools help me accomplish tasks more effectively".

Perceived Ease of Use (AI_EOU) similarly draws on established scales: "Learning to use GenAI tools is easy for me", "I find it easy to get GenAI tools to do what I want", "My interaction with GenAI tools is clear and understandable", "I find GenAI tools easy to integrate into my workflow".

AI Literacy (AI_Lit) requires multidimensional assessment: Technical understanding (how GenAI models work and their limitations), Critical evaluation skills (output verification, hallucination detection), Ethical reasoning (appropriate use contexts, bias awareness), and Practical application skills (prompt engineering, iteration strategies).

Actual Use (USE) can be measured through: Frequency (daily, weekly, monthly use), Breadth (number of task types and use cases), Depth (complexity of applications and integration level), and Self-reported and behavioral measures, where available.

Responsible Use (RUSE_I) at the individual level: Verification behaviors (checking outputs, cross-referencing), Appropriate task selection (avoiding prohibited uses), Bias awareness (considering potential biases in outputs), and Ethical compliance (adhering to organizational guidelines).

### E. *Outcome Measures*

Productivity and Task Performance (PROD) may be measured by analyzing Task completion time (relative to a pre-GenAI baseline), Output quality ratings (supervisor or peer assessment), Error rates (before and after GenAI adoption), and Self-reported productivity perceptions.

Individual Innovation (INNOV_I) can be quantified using Idea generation (number and quality of new ideas), Experimentation behavior (trying new approaches), Creative output (novel solutions, recombination), and Innovation self-efficacy.

Organizational Transformation (TRANSF) may be evaluated through Process redesign extent (workflows modified for GenAI), Role evolution (job descriptions, skill requirements), Capability development (new organizational competencies), and Structural adaptation (team configurations, reporting relationships).

Sustainable Competitive Advantage (SCA) can be measured through Relative performance (compared to industry peers), Agility indicators (speed of adaptation, responsiveness), Differentiation measures (unique capabilities, market position), and Sustained performance (longitudinal performance trajectories).

### F. *Research Design Considerations*

Testing the framework requires attention to several methodological considerations: Cross-level analysis, Temporal dynamics, Industry variation, and Measurement timing.

In what concerns cross-level analysis, the framework specifies relationships across organizational, workflow, and individual levels. Hierarchical linear modeling or multilevel structural equation modeling can accommodate this nested structure.

The relationships between investment, democratization, and outcomes unfold over time. Longitudinal designs capturing adoption trajectories and the evolution of outcomes are preferable to cross-sectional snapshots.

GenAI applicability and adoption patterns vary across industries. Multi-industry samples enable testing of boundary conditions and moderating effects.

Individual perceptions, use behaviors, and outcomes may operate on different time scales. Careful attention to measurement timing is required to capture hypothesized relationships.

## VII. Managerial Implications

The framework yields several actionable implications for managers seeking to realize competitive benefits from GenAI investments.

### A. *Treat GenAI as Augmentation Infrastructure*

Managers should conceptualize GenAI as infrastructure for workforce augmentation rather than solely as automation technology. This framing shifts attention from replacing human work to enhancing human capability. Augmenting infrastructure requires investment not only in technology but also in the human capabilities needed to leverage it effectively.

This perspective suggests different investment priorities than an automation-focused approach. Rather than concentrating GenAI deployment on narrow, high-volume tasks suitable for automation, augmentation infrastructure extends GenAI access broadly across the workforce, enabling employees to enhance diverse aspects of their work.

### B. *Expand Access with Literacy Development*

Expanding GenAI access without corresponding literacy development risks overreliance, misuse, and unrealized potential. Managers should pair access expansion with systematic literacy development, including: Foundational training, which ensures all employees understand how GenAI works, its limitations, and appropriate use contexts; Skill development, that encompasses building practical capabilities in prompt engineering, output verification, and workflow integration; Acceptable use guidelines, by establishing clear policies that enable experimentation while managing risks;

Verification habits consisting in cultivating organizational norms around checking AI outputs rather than accepting them uncritically.

The goal is not to restrict use but to enable responsible use that captures benefits while managing risks. As AI Literacy expanded in a systemic way, enables a collective capacity of the society, by including universities, institutions, organizations, and citizens, to critically understand and engage with AI technologies. This widespread literacy is essential for upholding ethical principles like fairness, transparency, and accountability, as it enables people to recognize and challenge practices such as ethics washing, lobbying, and responsibility shirking. In the context of AI democratization, systemic AI literacy acts as a societal safeguard, empowering communities to demand ethical AI development and hold developers accountable [13].

### C. *Measure Depth and Quality, Not Just Adoption*

Many organizations track GenAI adoption through license counts, login frequencies, or pilot program participation. These metrics capture access and initial use but miss the quality dimensions that drive value creation.

More meaningful measurement addresses: Access, Use quality, Task change, Capability growth.

Regarding access, we may ask: Who can use what, where, and in which workflow steps? The quality of use may be evaluated by asking whether employees verify outputs, understand the limits, and avoid overreliance. Task change: Which tasks become faster, better, or newly possible with GenAI? Capability growth: Does AI literacy improve experimentation, judgment, and innovation?

These quality-focused metrics provide better insight into whether GenAI investments are translating into organizational capability.

### D. *Use Governance to Enable Experimentation*

Governance frameworks should enable safe experimentation rather than creating purely restrictive regimes. Overly restrictive policies may protect against misuse risks but sacrifice the exploration and learning required for effective democratization. Effective governance balances: Clarity consists of clear guidelines on acceptable and prohibited uses; Flexibility involves making room for experimentation within defined boundaries; Learning captures insights from employee experimentation; Adaptation updates policies as understanding evolves. [8]

Organizations that achieve this balance can realize innovation benefits while managing risks, creating governance capability that competitors may find difficult to replicate.

## VIII. 8. Contributions and Future Research

### A. *Theoretical Contributions*

This paper makes three primary contributions to the literature on AI and competitive advantage.

First, we shift the analytical lens from firm-level adoption to employee level value creation. While prior research has examined organizational AI adoption and its performance implications, less attention has focused on the micro foundations through which AI investments translate into competitive benefits. By centering individual level AI democratization, we address calls for micro level theorizing in information systems research and provide a more granular understanding of the adoption and performance relationship.

Second, we build a cross-level conceptual model with explicit micro foundations and testable propositions. The framework specifies constructs at organizational, workflow, and individual levels, articulates cross-level mechanisms linking these constructs, and advances propositions that can guide empirical research. This multi-level architecture enables researchers to examine how organizational investments shape individual behaviors and how individual outcomes aggregate to organizational performance.

Third, we provide a measurement scaffold for empirical work across industries and over time. The operationalization section offers illustrative indicators for each construct, enabling researchers to develop measurement instruments appropriate to their research contexts. This scaffold supports cumulative empirical research that can test, refine, and extend the framework.

### B. *Practical Contributions*

For practitioners, the framework offers guidance on realizing competitive benefits from GenAI investments. The emphasis on democratization quality rather than mere adoption directs attention to literacy development, the cultivation of responsible use, and the design of governance. The measurement recommendations enable organizations to assess their progress toward democratization and identify opportunities for improvement.

### C. *Limitations*

Several limitations should be acknowledged. The framework is conceptual; empirical testing is required to validate the proposed relationships. The propositions are directional but do not specify effect sizes or boundary conditions that may moderate relationships. The framework focuses on individual-level democratization but does not fully elaborate on team-level dynamics that may influence adoption and use patterns.

### D. *Future Research Directions*

We identify four priority directions for future research.

Longitudinal studies of democratization dynamics: Cross-sectional research cannot capture how democratization unfolds over time. Longitudinal designs tracking adoption trajectories, literacy development, and outcome evolution would illuminate the temporal dynamics of AI democratization and identify critical periods or tipping points in the adoption process [8].

Cross-cultural investigations: AI adoption patterns, literacy development, and responsible use norms may vary across cultural contexts. Cross-cultural research can identify universal mechanisms and culturally contingent factors, informing both theory and practice in global organizations[17].

Industry-specific boundary conditions: GenAI applicability varies across industries based on task characteristics, regulatory environments, and competitive dynamics. Industry-focused research can identify boundary conditions that moderate framework relationships and develop industry-specific guidance[12].

Governance and transformation dynamics: The relationship between governance approaches and transformation outcomes warrants deeper investigation [18] using several research approaches [2], [18], [19], [20], [21]. Research examining how different governance configurations, restrictive vs. enabling, centralized vs. distributed influence democratization quality and organizational transformation would inform governance design.

Team level mechanisms: While our framework focuses on individual and organizational levels, team dynamics likely play important roles in shaping adoption patterns, use norms, and outcome aggregation. Research examining team level mechanisms would enrich the framework's multi-level architecture [17].

## IX. Conclusions

As generative AI diffuses and becomes broadly accessible, sustainable competitive advantage increasingly depends on workforce wide AI democratization. Organizations that invest in GenAI but fail to achieve broad, responsible, effective employee adoption will not realize competitive benefits. Conversely, organizations that cultivate individual level AI democratization grounded in perceived usefulness, ease of use, and AI literacy can sustain advantage even as technology commoditizes.

This paper has developed a conceptual framework linking firm level GenAI investment and governance to sustainable competitive advantage through individual level AI democratization. The framework specifies constructs at multiple levels, articulates cross-level mechanisms, and advances testable propositions. We have provided operationalization guidance for empirical research and derived managerial implications for practitioners.

The democratization of generative AI represents both opportunity and challenge. Organizations that treat GenAI as augmentation infrastructure, invest in workforce literacy, measure use quality rather than mere adoption, and design governance for experimentation can build organizational capabilities that competitors cannot easily replicate. In a world where everyone has access to powerful AI tools, the organizations that thrive will be those that empower their people to use those tools meaningfully, responsibly, and effectively.